\documentclass[twocolumn]{revtex4}
\usepackage{graphicx, verbatim}

\begin{document}

\title{Comment on arXiv:1807.08572, 
	``Coexistence of Diamagnetism and Vanishingly Small Electrical Resistance at
	Ambient Temperature and Pressure in Nanostructures"}

\author{Sergei Urazhdin}
\affiliation{Department of Physics, Emory University,
Atlanta, GA 30322}

\begin{abstract}

A recent preprint arXiv:1807.08572 reported the observation of a transition in Ag/Au nanoparticle composites near room temperature and at ambient pressure, to a vanishingly small four-probe resistance, which was tentatively identified as a percolating superconducting transition. In this brief comment, I point out that a vanishing four-probe resistance may also emerge in non-superconducting systems near conductance percolation threshold.

\end{abstract}
\maketitle

Article arXiv:1807.08572, which was originally posted in 2018 under the title ``Evidence for Superconductivity at Ambient Temperature and
Pressure in Nanostructures", teased the scientific community with a tantalizing prospect of achieving superconductivity at ambient conditions, and generated a significant media attention~\cite{nature}. The extraordinary claims attracted a close scrutiny of the reported data. In particular, repeated noise pattern identified in one the Figures in the article was pointed out as evidence against the credibility of the results~\cite{Skinner}. Since then, the authors posted a revised manuscript with a modified title ``Coexistence of Diamagnetism and Vanishingly Small Electrical Resistance at Ambient Temperature and Pressure in Nanostructures", providing more detailed information about their sample preparation and measurements, as well as multiple additional datasets demonstrating the reproducibility of resistive transition in a substantial fraction of their samples. Furthermore, a recently posted follow-up article arXiv:1906.02291 "Current-voltage characteristics in Ag/Au nanostructures at resistive transitions" described additional measurements that leave little doubt about the existence of a resistive transition in the studied Ag/Au composites.

\begin{figure}
	\includegraphics[width=3.0in]{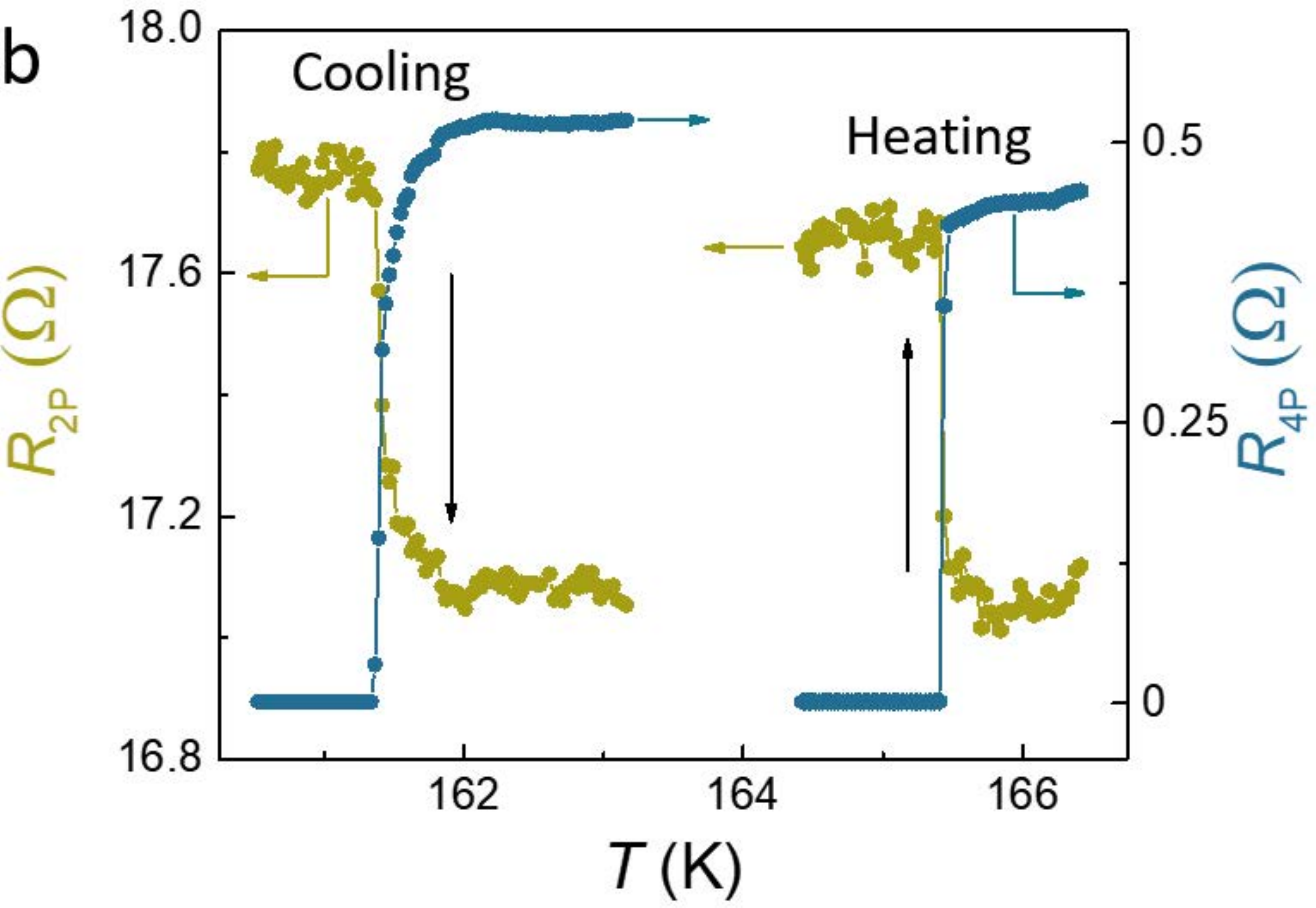}
	\caption{\label{fig:fig1} Figure 4(b) of Ref.~\cite{Islam}.}
\end{figure}

\begin{figure}
	\includegraphics[width=3.2in]{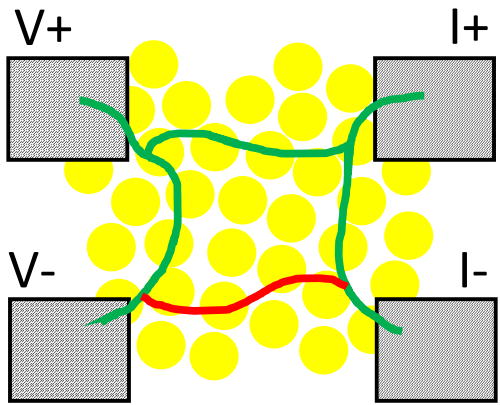}
	\caption{\label{fig:fig2} Schematic showing how a zero voltage can be observed in a four-probe measurement for a system with percolating conductance. When all three paths shown with green, and the path shown with red are connected, the four-probe resistance is finite. However, when the red path is disconnected, the four-probe resistance becomes zero, even though every pair of electrodes remain connected by a conducting path.}
\end{figure}

In my view, it is essential for the progress of science to look for new phenomena, often in unexpected places. However, it is equally important to critically examine the new theories and experimental observations. In this brief communication, I point out that the observation of zero voltage in a four-probe measurement does not necessarily imply the presence of superconductivity, and may be instead associated with other, less exotic phenomena. This possibility is suggested by Figure 4(b)  of Ref.~\cite{Islam}, reproduced as Fig.~\ref{fig:fig1} in this communication, which shows both the four- and the two-probe measurements performed on heating and cooling of one of the samples. The four-probe resistance $R_{4P}$, defined as the ratio of the voltage measured across one pair of electrodes to the current across another pair of electrodes, drops to zero below a certain temperature. In contrast, the two-probe resistance $R_{2P}$, defined as the ratio of the voltage to current across the same pair of electrodes, exhibits an abrupt increase in the same temperature range. If the resistive transition is indeed associated with the onset of superconductivity, one would generally expect to observe a decrease of $R_{2P}$ to a finite value determined by the contact and the lead resistance. The abrupt increase of $R_{2P}$ suggests that the studied composite film may be close to the conductivity percolation threshold. The existence and the properties of the conductive paths between a pair of electrodes may be affected by perturbations such as thermal stress, voltage/current, and magnetic field. If some of the conducting paths between the current leads abruptly open, the two-probe resistance increases. In contrast, in the four-probe measurement, breaking of the conducting path can result in a vanishing voltage across the voltage electrodes. In the simplest scenario, this would occur if at least one of the voltage electrodes becomes electrically isolated from the current path. However, this is also possible with all the electrodes remaining connected, as illustrated in Fig.~\ref{fig:fig2}, such that there is no contradiction with the metallic two-electrode behaviors shown in Fig.1[d] of  Ref.~\cite{Islam}. 

Mechanical breaking of conductive paths near the percolation threshold, induced by thermal stress, may provide a reasonable explanation for the hysteresis in the temperature- and current-dependences reported in Refs.~\cite{Thapa} and ~\cite{Islam}, variations observed in thermal cycling, re-entrant transitions, and incomplete resistive switching where the four-probe resistance does not drop to zero. Nevertheless, other evidence presented in these articles (such as onset of diamagnetism concurrent with resistive transition) is harder to reconcile with such a simple explanation, which in my opinion certainly warrants further studies by a variety of advanced techniques available to modern scientific research.

\begin{thebibliography}{99}
\bibitem{Thapa} 
D. K. Thapa et al., Coexistence of Diamagnetism and Vanishingly Small Electrical Resistance at
Ambient Temperature and Pressure in Nanostructures, arXiv:1807.08572 (2018).
\bibitem{nature} D. Castelvecchi, Physicists doubt bold superconductivity claim following social-media storm, Nature, (2018).
\bibitem{Skinner} B. Skinner, Repeated noise pattern in the data of arXiv:1807.08572, ``Evidence for Superconductivity at Ambient Temperature and
Pressure in Nanostructures", arXiv:1808.02929 (2018).
\bibitem{Islam} S. Islam et al., Current-voltage characteristics in Ag/Au nanostructures at resistive transitions, arXiv:1906.02291 (2019).

\end{thebibliography}
\end{document}